\documentclass{PoS}

\usepackage{amssymb}
\usepackage{bm}
\usepackage{amsthm}
\usepackage{xcolor}
\usepackage{amsmath}
\usepackage{graphicx}

\newcommand{\CC}{\mathbb{C}}
\newcommand{\RR}{\mathbb{R}}

\newcommand{\NN}{\mathbb{N}}

\newcommand{\Hi}{{\cal H}}
\newcommand{\Ob}{{\cal O}}

\newcommand{\A}{{\cal A}}

\newcommand{\R}{{\cal R}}
\newcommand{\bv}{{\bf v}}
\newcommand{\bx}{{\bf x}}

\newcommand{\bpi}{{\bm\pi}}

\newcommand{\be}{\begin{equation}}
\newcommand{\ee}{\end{equation}}
\newcommand{\bea}{\begin{eqnarray}}
\newcommand{\eea}{\end{eqnarray}}
\newcommand{\ba}{\begin{array}}
\newcommand{\ea}{\end{array}}

\title{New fuzzy spheres through confining potentials and energy cutoffs}

\ShortTitle{New fuzzy spheres through confining potentials and energy cutoffs}

\author{\speaker{Gaetano Fiore},  Francesco Pisacane
\\
       Dip. di Matematica e Applicazioni, Universit\`a di Napoli ``Federico II'',\\
\& INFN, Sezione di Napoli, \\
Complesso Universitario  M. S. Angelo, Via Cintia, 80126 Napoli, Italy\\
        E-mail: \email{gaetano.fiore@unina.it},  \email{francesco.pisacane@unina.it}}

\abstract{We briefly report on our recent construction \cite{FioPis17} of new fuzzy spheres $S^d_{\Lambda}$ of dimensions $d=1,2$ covariant under the {\it full} orthogonal group $O(D)$, $D=d\!+\!1$.  $S^d_{\Lambda}$  is built imposing a suitable energy cutoff on a quantum particle in $\RR^D$ subject to a confining potential well $V(r)$  
with a very sharp minimum on the sphere of radius $r=1$; the cutoff and the depth of the well depend
on (and diverge with) $\Lambda\in\NN$.
The commutator of the coordinates depends only on the angular momentum, as in
 {\it Snyder} noncommutative spaces. As  $\Lambda\to\infty$ the Hilbert
space dimension diverges, $S^d_{\Lambda}\to S^d$, 
and we recover ordinary quantum mechanics on $S^d$.  
These models might be useful in quantum field theory,  quantum gravity or condensed matter physics.}

\FullConference{Corfu Summer Institute 2017 "School and Workshops on Elementary
Particle Physics and Gravity", \ 
                 2-28 September 2017, \
                 Corfu, \ Greece}

\begin{document}

\section{Introduction}

Nowadays noncommutative space(time) algebras are introduced and studied
for several reasons.  In particular: \ to regularize ultraviolet (UV) divergences in quantum field theory
(QFT) \cite{Snyder,Kad62,Mir67}; 
\ as an arena to formulate Quantum Gravity (QG) that naturally induces 
uncertainty relations of the type $\Delta x\gtrsim L_p$, as predicted by QG arguments (see e.g.
\cite{Wee57,Des57,Mea64,DopFreRob95}); 
\ as an arena for the unification of fundamental interactions  (see e.g. \cite{ConLot90,ChaCon10}).
Today  Noncommutative Geometry  \cite{Connes,Madore99,GraFigVar00,Lan97} 
is a sophisticated framework that  develops the whole machinery of differential geometry  on
noncommutative spaces. 
Fuzzy spaces are particularly appealing noncommutative spaces: a fuzzy space is a sequence $\{\mathcal{A}\}_{n\in\NN}$
of {\it finite-dimensional} algebras such that
 $\mathcal{A}_n\overset{n\rightarrow\infty}\longrightarrow\mathcal{A}\!\equiv$algebra 
of regular functions on an ordinary manifold, with \ $dim(\mathcal{A}_n)\overset{n\rightarrow\infty}\longrightarrow\infty$. 
The first and seminal fuzzy space is the Fuzzy Sphere (FS) of Madore and Hoppe \cite{Mad92,HopdeWNic},
the first applications to QFT
models are in \cite{GroMad92,GroKliPre96'}; 
${\cal A}_n\simeq M_n(\CC)$ is generated by coordinates $x^i$ ($i=1,2,3$) fulfilling
\be
[x^i,x^j]=\frac {2i}{\sqrt{n^2\!-\!1}}\varepsilon^{ijk}x^k, \quad
r^2:=x^ix^i=1,\qquad n\in\NN\setminus \{1\}                      \label{FS}
\ee
(sum over repeated indices is understood);
in fact they are obtained by the rescaling $x^i=2L_i/{\sqrt{n^2\!-\!1}}$ of the elements
$L_i$ of  the standard basis of $so(3)$ in the 
irreducible representation $(\pi_l,V_l)$ characterized by $L_iL_i=l(l+1)$,  $n=2l\!+\!1$.
Fuzzy spaces have raised a big interest in the high energy physics community as a non-perturbative technique in  QFT
 based on a finite-discretization of space(time)  alternative to the lattice one: the main advantage is that the algebras 
$\A_n$ can carry representations of Lie groups (not only of discrete ones). They can be used also 
 for  internal (e.g. gauge) degrees of freedom (see e.g. \cite{AscMadManSteZou}),
or  as a new tool in string and $D$-brane theories (see e.g. \cite{AleRecSch99,HikNozSug01}).

Relations (\ref{FS}) are covariant under $SO(3)$, but not under the whole $O(3)$; 
in particular {\it not under parity $x^i\mapsto -x^i$}, as the ordinary sphere $S^2$. In our opinion, another 
reason why the FS does not approximate  $S^2$ in the best possible way is that $V_l$
carries an {\it irreducible} representation of $SO(3)$ (so that the Casimir $r^2$ is
identically 1), whereas the Hilbert space of a quantum particle on $S^2$ has the following
decomposition in irreducible representations of $SO(3)$:
\be
{\cal L}^2(S^2)=\bigoplus\limits_{l=0}^\infty V_l.
\label{directsum}
\ee

Here we present new fuzzy approximations \cite{FioPis17} of quantum mechanics (QM)
on  $S^d$ ($d=1,2$) overcoming these two problems:
We start with an ordinary zero-spin quantum particle in $\RR^D$  configuration space ($D=d\!+\!1$) 
with Hamiltonian
\bea
H=-\frac 12\Delta + V(r).                                                 \label{Ham}
\eea
Here $r^2:= x^ix^i$, $\Delta:= \partial_i\partial_i$, 
$\partial_i\equiv \partial/\partial x^i$, $i=1,...,D$; we use dimensionless cartesian coordinates, momentum components and Hamiltonian \ $x^i,p_i:=-i\partial_i,H$. \ $x^i,p_i$ generate the Heisenberg algebra $\Ob$ of observables.
The  canonical commutation relations \
$[x^i,x^j]=0,\:\: [p_i,p_j]=0,\:\: [x^i,p_j]= i\delta^i_j $ \
as well as the Hamiltonian are invariant under all orthogonal transformations 
$x^i\mapsto x'{}^i=Q^i_j x^j$ ($Q^{-1}=Q^T$),    
including parity $Q=-I$. 
We choose $V(r)$ as a confining potential  with a very sharp minimum at \ $r=1$, i.e. with $V'(1)=0$ and
very large $k:= V''(1)/4>0$, \ and fix 
$V_0:= V(1)$ so that the ground state has zero energy,  $E_0=0$. We  choose an energy cutoff $\overline{E}$
satisfying first of all the condition
\be
V(r)\simeq V_0+2k (r-1)^2\qquad \mbox{if $r$ fulfills}\quad V(r)\le  \overline{E},
\label{cond1}
\ee
so that $V(r)$ is approximately  harmonic  in the
classical region $v_{\overline{E}}$ compatible with the energy cutoff \ $V(r)\le \overline{E}$. \ 
Then we project the theory onto the finite-dimensional Hilbert subspace $\mathcal{H}_{\overline{E}}\subset\Hi\equiv \mathcal{L}^2(\RR^D)$ 
spanned by $\psi$ fulfilling the eigenvalue equation
\be
H\psi=E\psi, \qquad\psi\in {\cal L}^2\left(\mathbb{R}^D\right),                      \label{Heigen}
\ee
with  $E\leq\overline{E}$.
This entails  replacing every observable $A$ by $\overline{A}$: 
$$
A\mapsto\overline{A}:=P_{\overline{E}}AP_{\overline{E}},
$$
where $P_{\overline{E}}$
is the projection on $\mathcal{H}_{\overline{E}}$. 
$H,L_{ij},P_{\overline{E}}$ commute;  $L_{ij}:=x^ip_j-x^jp_i$ are the angular momentum components.
Decomposing the Laplacian on $\RR^D$ 
in polar coordinates $r,\varphi,...$ 
\be
\Delta=\partial_r^2+(D-1)\frac 1r\partial_r-\frac 1{2r^2}L_{ij}L_{ij},  \label{LaplacianD}
\ee
recalling that the eigenvalues of the square angular momentum $L^2=L_{ij}L_{ij}/2$ are $j\left(j+D-2\right)$,
 and using the Ansatz $\psi=f(r)Y(\varphi,...)$  \ [$Y$ are eigenfunctions of $L^2$ and of the elements of a Cartan subalgebra of $so(D)$] we transform (\ref{Heigen})  into this auxiliary ODE in the unknown $f(r)$:
\be
\left[-\partial_r^2-\frac {D-1}r\partial_r+\frac {j\left(j+D-2\right)}{r^2}+V(r)\right]f(r)=Ef(r).\label{eqpolar}
\ee
To obtain the lowest eigenvalues at leading order in $1/k$
we don't need to solve it exactly: condition (\ref{cond1}) allows us to approximate (\ref{eqpolar}) with the eigenvalue equation
 of a $1-$dimensional harmonic oscillator, by Taylor expanding \ $V(r)$, $1/r$, $1/r^2$ \
around $r=1$.

As a second condition on the cutoff we ask that it be
{\it sufficiently low to  `freeze' radial excitations}, so that the eigenvalues of $H$ fulfilling
$E\le \overline{E}$ coincide at leading order
with those of  the square angular momentum  $L^2=L_{ij}L_{ij}$, i.e.
with the Laplacian on the sphere $\mathbb{S}^d$; this can be considered as a quantum version of the constraint $r=1$.   
It turns out that
on $\Hi_{\overline{E}}$ the $x^i$ are noncommutative {\it \`a la Snyder}, namely
their commutators depend only on the angular momentum, and that  
they generate the whole algebra of observables.  The whole procedure
is $O(D)$-covariant by construction. Making $\overline{E}$,  $V''(1) \gg 0$ diverge with  
some $\Lambda\!\in\!\NN$ (while $E_0\!=\!0$), 
and keeping the leading terms in $1/\Lambda$,
we get a sequence $\{\A\}_{\Lambda\in\NN}$ of fuzzy approximations of ordinary 
quantum mechanics (QM) on $S^d$. 
 On  $\Hi_{\overline{E}}\equiv\Hi_{\Lambda}=\bigoplus_{l=0}^\Lambda V_l$ the square distance $\mathcal{R}^2$ from the origin is not identically 1, but
a function of $L^2$, whose spectrum collapses to 1 in the $\Lambda\to\infty$  limit.

\medskip
Our construction is inspired by the Landau model, where noncommuting $x,y$  are obtained
projecting QM with a strong uniform magnetic field $B$ on the lowest energy subspace; therefore 
the method  is {\it physically sound}. 
Our models might have applications to quantum models 
in condensed matter physics with an effective
 one- or two-dimensional configuration space in the form of a circle, a cylinder
 or a sphere,
because  they respect parity, and  the restriction to the circle, cylinder or sphere is an effective one obtained ``a posteriori'' from the exact dynamics in the physical dimension 3.
But we think that they are interesting also as 
new toy-models of fuzzy geometries in quantum field theory,  quantum gravity, string theory. Our procedure can  be generalized in a straightforward manner
to $D>3$, as well as to other confining potentials; 
the dimension of the accessible Hilbert space $\Hi_{\overline{E}}$
 will be approximately ${\cal B}/h^D$, 
where  $h,{\cal B}$ are the Planck constant and the  volume of the 
classically allowed region in phase space
(i.e. the one characterized by energies $E\le \overline{E}$). 
If $H$ is invariant under some symmetry group, then the projection $P_ {\overline{E}}$
on  $\Hi_{\overline{E}}$ is invariant as well, and the projected theory will inherit that symmetry.
 Imposing a cutoff $\overline{E}$ on a given theory may have various motivations,
in particular: it can yield an effective description of a system when our preparation
of the system, or our measurements, or  the interactions with the environment, 
cannot bring its state to energies $E>\overline{E}$; or it may even be a necessity if we believe
$\overline{E}$  represents the threshold for the onset of new physics  not
accountable by that theory.

In sections \ref{D2}, \ref{D3} we treat 
the cases $D=2,3$ leading to $S^1_\Lambda,S^2_\Lambda$ respectively. 
Section \ref{Conclu} contains a comparison with the literature,
final remarks, outlook and
conclusions. For more details, explicit computations and  proofs we refer the reader to 
Ref.  \cite{FioPis17}.

  \begin{figure}
        \begin{minipage}[c]{.48\textwidth}
          \includegraphics[scale=0.19]{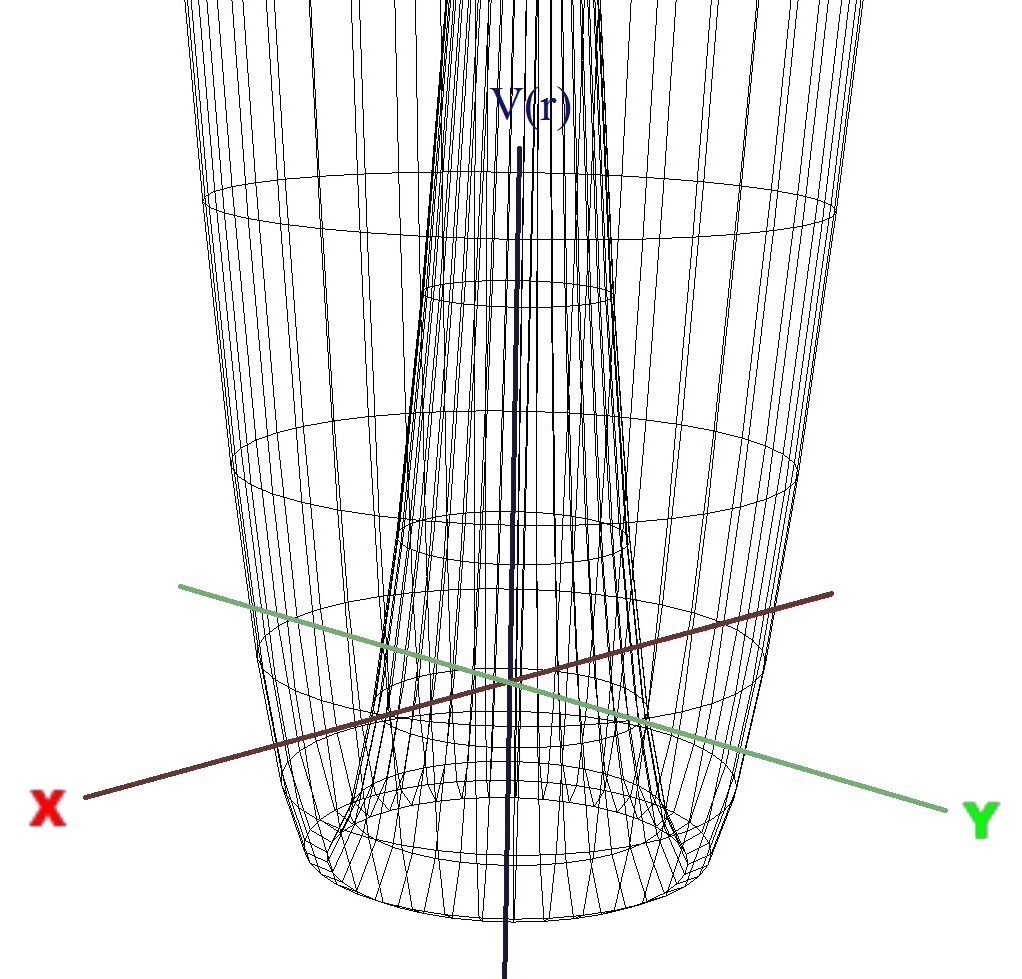}
          \caption{Three-dimensional plot of $V(r)$}
        \end{minipage}%
        \hspace{5mm}%
        \begin{minipage}[c]{.48\textwidth}
\begin{center}          \includegraphics[scale=0.35]{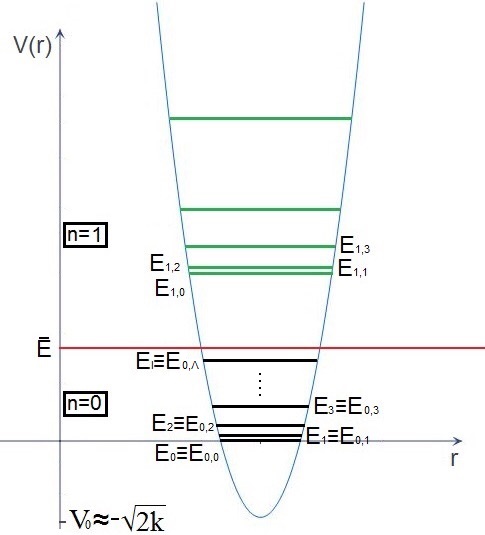}
          \caption{Two-dimensional plot of $V(r)$ including the energy-cutoff}
\end{center}        \end{minipage}
\label{fig1}
      \end{figure}

\section{$D=2$: $O\left(2\right)$-covariant fuzzy circle}
\label{D2}

The  potential is shown in fig. 1. 
For convenience we look for the solution $\psi$
of (\ref{Heigen})  in the form \ $\psi=e^{i m\varphi}f(\rho)$,  
\ with \ $m\in\mathbb{Z}\equiv$spectrum of $L\equiv L_{12}$, \  $\rho:=\ln r$, \
and expand around $\rho=0$.  \ 
The harmonic oscillator approximation of (\ref{eqpolar}) 
has eigenvalues and (H\'ermite) eigenfunctions 
\bea
&& E=E_{n,m}={\color{red}{2n\sqrt{\!2k}-2n}}+{\color{blue}{m^2}} +O\!\left(1/\sqrt k\right) \label{Enm}\\[8pt]
&& \ba{l}
f_{n,m}(\rho)=N_{n,m}\exp\left[\!-\frac{\left(\rho\!-\!\widetilde{\rho}_{n,m}\right)^2\!\sqrt{\! k_{n,m}}}{2}\right] H_n\!\left[\left(\rho\!-\!\widetilde{\rho}_{n,m}\right)\sqrt[4]{\! k_{n,m}}\right],\\[8pt]
k_{n,m}=2(k\!-\!E_{n,m}\!+\!V_0), \qquad \widetilde{\rho}_{n,m}=\frac{E_{n,m}\!-\!V_0}{k_{n,m}},
\ea\qquad 
\eea
with \ $n\in\mathbb{N}_0$; we have set $V_0=-\sqrt{2k}\!+\!2\!+\!O\!\left(\frac{1}{\sqrt k}\right)$ to fulfill
the requirement $E_{0,0}=0$. 
Up to terms $O\!\left(\frac{1}{\sqrt k}\right)$ (\ref{Enm}) gives
\be
E_m\equiv E_{0,m}={\color{blue}{m^2}},
\ee
which are the eigenvalues of the Laplacian  $L^2$ on $S^1$, while $E_{n,m}\to\infty$ as $k\!\to\!\infty$
if $n\!>\!0$ we have (we have highlighted the unwelcome $n$-dependent contribution  in red,  the welcome $m$-dependent one in blue); 
we can eliminate the latter eigenvalues by choosing
a cutoff \ $\overline{E} <2\sqrt{\!2k}\!-\!2$. \
The eigenfunctions of $H$  corresponding to $E=E_m$ are
$$
\psi_m\left(\rho,\varphi\right)=N_me^{im\varphi} 
e^{-\frac{\left(\rho\!-\!\widetilde{\rho}_m\right)^2\sqrt{k_m}}{2}}.
$$
We set \ ${\Lambda}\!:=\!\left[\sqrt{\overline{E}}\right]\in\NN$ and abbreviate $\mathcal{H}_{\Lambda}\!\equiv\!\mathcal{H}_{\overline{E}}$. \ $E_m\!\le\!\overline E$ implies   
\begin{equation}
m^2\leq\Lambda^2 <2\sqrt{\!2k}-2          \label{consistency}
\end{equation}
so that all $E_m$ are smaller than the energy levels corresponding to $n>0$ (see fig. 2).
Clearly dim$(\Hi_{\Lambda})\!=\!2\Lambda\!+\!1$. 
We recover the whole spectrum of $L^2$ on $S^1$ by allowing $\sqrt{\overline{E}}$, or equivalently
$\Lambda$, to diverge with $k$ while respecting (\ref{consistency}).

Let $x^{\pm}:=\frac{x\pm iy}{\sqrt{2}}=re^{\pm i\varphi}$. 
By explicit computations
\begin{equation}
\langle\psi_n, x^{\pm}\psi_m\rangle= \frac{a}{\sqrt{2}}\left[1+\frac{m(m\pm1)}{2k}\right] \delta^n_{m\pm1} \label{psimx}
\end{equation}
with $a=1\!+\!\frac 94\frac{1}{\sqrt{2k}}\!+\!\frac{137}{64k}\!+\!...$. To get rid of $a$ we rescale 
$\xi^\pm:=\frac{\overline{x}^\pm}a$. 
$\overline{x}^{\scriptscriptstyle -}\!,\xi^{\scriptscriptstyle -}$ are resp. the adjoints of $\overline{x}^{\scriptscriptstyle +}\!,\xi^{\scriptscriptstyle +}$. Then,
 up to terms $O(1\!/\!k^{3/2})$
\bea
\ba{lll} \xi^{\pm}\psi_m &=&
\left\{\!\!\begin{array}{ll}\!\!
  \frac{1}{\sqrt{2}}\left[1+\frac{m(m\pm1)}{2k}\right]\psi_{m\pm1} \:\:&
\mbox{if }-\!{{\Lambda}}\leq \pm m\leq{{\Lambda}}\!-\!1 \\[8pt]
0 & \mbox{otherwise,}
\end{array}\right.          \\[20pt]
\overline{L}\,\psi_m &=& m\,\psi_m.
\ea\label{bary+}
\eea
We define the square distance from the origin as the
$O(2)$-invariant \  $\mathcal{R}^2:=\xi^+\xi^-+\xi^-\xi^+$; let $\widetilde{P}_m$ be the projection over the $1$-dim subspace spanned by $\psi_m$. Eq. (\ref{bary+}) implies at leading order in $1/k$
\begin{equation}
\left[\xi^+,\xi^-\right]=-{\color{blue}\underbrace{\frac{\overline{L}}k}_{Snyder-like}}+\left[1\!+\!\frac {{{\Lambda}}({{\Lambda}}\!+\!1)}k\right]\!\frac{\widetilde P_{{{\Lambda}}}\!-\!\widetilde P_{-{{\Lambda}}}}2.\label{y+y-}
\end{equation}
\begin{equation}
\prod\limits_{m=-{{\Lambda}}}^{{{\Lambda}}}\!\!\left(\overline{L}\!-\!mI\right)=0, \qquad \left(\overline{L}\right)^\dagger=\overline{L},\label{commrelD=2}
\end{equation}
\begin{equation}
\left[\overline{L}, \xi^{\pm}\right]=\pm \xi^\pm,\quad \xi^+{}^\dagger=\xi^-, \quad\left(\xi^\pm\right)^{2{{\Lambda}}+1}=0.\label{commrelD=2'}
\end{equation}
\begin{equation}
\mathcal{R}^2=  1+\frac{\overline{L}^2}{k} -
\left[1\!+\!\frac {{{\Lambda}}({{\Lambda}}\!+\!1)}{k}\right]
\frac{\widetilde P_{{{\Lambda}}}\!+\!\widetilde P_{-{{\Lambda}}}}2.          \label{defR2D=2}
\end{equation}
Eq. (\ref{y+y-}-\ref{defR2D=2}) are {\it exact} if we adopt (\ref{bary+}) as {\it definitions} of 
$\xi^+,\xi^-,\overline{L}$. 
To obtain a fuzzy space we can 
choose $k$ as a function of ${{\Lambda}}$ fulfilling (\ref{consistency}), for example $k={{\Lambda}}^2({{\Lambda}}\!+\!1)^2$, and the commutative
limit will be ${{\Lambda}}\to \infty$.
Then e.g. (\ref{y+y-}) becomes
\begin{equation}
[\xi^+,\xi^-]=\frac{-\overline{L}}{ {{\Lambda}}^2({{\Lambda}}\!+\!1)^2}+
\left[1\!+\!\frac 1{ {{\Lambda}}({{\Lambda}}\!+\!1)}\right]\!
\frac{\widetilde P_{{{\Lambda}}}\!-\!\widetilde P_{-{{\Lambda}}}}2.           \label{comrelD=2'}
\end{equation}

Let us summarize what we have found so far:
\begin{itemize}

\item The matched {\it confining potential} and  {\it energy cutoff}
lead to a non-zero commutator of the coordinates of the Snyder's Lie algebra type, i.e. depending only on  $L$,
and vanishing as $k\to\infty$. To obtain a fuzzy space we can choose $k$ as a function of $\Lambda$ fulfilling (\ref{consistency}); and the commutative limit will be $\Lambda\rightarrow+\infty$.

\item $\mathcal{R}^2\neq 1$, but it is a function of $L^2$; its eigenvalues  
(except on $\pm\Lambda$)
are close to $1$, slightly grow with $|m|$ and collapse to 1 as $\Lambda\to \infty$.

\item Relations (\ref{y+y-}-\ref{defR2D=2})  are $O(2)$-invariant,  because in the  original model
both the commutation relations and  $H$ (hence also  $P_{\overline E}$) are. 

\item The ordered monomials  $(\xi^+)^h(\overline{L})^l(\xi^-)^n$ [with degrees $h,l,n$
bounded by (\ref{commrelD=2}-\ref{commrelD=2'})] make up a 
 basis of the $*$-algebra of observables \ $\A_\Lambda\!:=\!End(\Hi_\Lambda\!)$ [a $(2\Lambda\!+\!1)^2$-dim  vector space]; for instance, the $\widetilde{P}_m$ themselves can be expressed as polynomials in $\overline{L}$. 

\item $\xi^+\!,\xi^-$ (or equivalently $\overline{x}^+\!,\overline{x}^-$) generate $\A_\Lambda$,
because also $\overline{L}$ can be expressed as a non-ordered polynomial in $\xi^+\!,\xi^-$. 
An alternative set of generators is $\{E^+,E^-\}$ in the $(2\Lambda\!+\!1)$-dimensional 
representation of $Usu(2)$ (see below).

\item As $\Lambda\to \infty$ \ $[\xi^+,\xi^-]\!\to\!0$,  
dim$(\Hi_{\Lambda})\!\to\!\infty$,  
$\psi_m\to\delta(\rho) e^{im\varphi}$.

\end{itemize}

What about the operators  $\overline{\partial}_\pm$?   
As seen, they are not needed as generators of $\A_\Lambda$. Actually, 
$\overline{\partial}_\pm$ do not go to $\partial_\pm$  as $\Lambda\to \infty$ 
because every $\partial_\pm\psi_m$ has a non-negligible $n=1$ component.
On the contrary, $\overline{L}\to L$; this is welcome, because in the limit $\Lambda\to \infty$
 all vector fields tangential to $S^1$ are of the form \ $f(\varphi) L$.

\subsection{Realization of the algebra of observables through $Uso(3)$}

The algebra of observables $\A_{\Lambda}:=End(\Hi_{{{\Lambda}}})$ is isomorphic to
 \be
\A_{\Lambda}\simeq M_N(\CC)\simeq\pi_\Lambda[Uso(3)], \qquad 
N=2\Lambda\!+\!1,                                 \label{isomD2}
\ee
where $\pi_{{{\Lambda}}}$ is the $N$-dimensional unitary representation  of $Uso(3)$.
The latter is characterized by the condition $\pi_{{{\Lambda}}}(C)={{\Lambda}}({{\Lambda}}+1)$, where  $C=E^aE^{-a}$ is the Casimir, and $E^a$ ($a\in\{+,0,-\}$)
make up the Cartan-Weyl basis $E^a$ of $so(3)$,
\begin{equation}
[E^+,E^-]=E^0,\qquad [E^0,E^\pm]=\pm E^\pm,\qquad E^a{}^\dagger=E^{-a}.
 \label{su2rel}
\end{equation}
To simplify the notation we drop $\pi_{{{\Lambda}}}$.
In fact we can realize \  $\xi^+,\overline{L}\, , \xi^-$ \ by setting
\begin{equation}
\ba{c}
\overline{L}=E^0, \qquad  \xi^\pm=f_{\pm}(E^0)E^\pm,\\[10pt]
\displaystyle\mbox{where}\quad  f_{+}(s)=\,\sqrt{\frac{1\!+\!s(s\!-\!1)/k}{{{\Lambda}}({{\Lambda}}+1)\!-\!s(s\!-\!1)}}= f_{-}(s-1).
\ea\label{transfD2}\
\end{equation}

\subsection{$*$-Automorphisms of the algebra of observables}

Within the group $SU(N)$ of  $*$-automorphisms of $M_N(\CC)\simeq \A_{\Lambda}$ 
\be
a\mapsto g\, a \, g^{-1}, \qquad a\in \A_{\Lambda}\simeq M_N, \quad g\in SU(N),  \label{autom}
\ee
a special role is played 
by the subgroup $SO(3)$ acting through the representation $\pi_\Lambda$, namely $g=\pi_\Lambda\left[e^{i\alpha}\right]$,
where $\alpha\in so(3)$  is a combination with real coefficients of \ $E^0, E^+\!+\!E^-,i(E^-\!-\!E^+)$.
$O(2)\subset SO(3)$ plays the role of isometry group. \ In particular, choosing $\alpha=\theta E^0$ amounts
to a rotation by an angle $\theta$ in the  $\overline{x}^1\overline{x}^2$ plane:  $\overline{L}\mapsto \overline{L}$ and
$$
\overline x^\pm\mapsto \overline x'{}^\pm = e^{\pm i\theta}\overline x^\pm \qquad
\Leftrightarrow\qquad\left\{ \begin{array}{l } \overline x'{}^1=\overline x{}^1\cos\theta+\overline x{}^2\sin\theta  \\ \overline x'{}^2=- \overline x{}^1\sin\theta +\overline x{}^2\cos\theta \end{array} \right. .
$$ 
Choosing \ $\alpha=\pi (E^+\!+\!E^-)/\sqrt{2}$ 
we obtain a  $O(2)$-transformation
with determinant $=-1$ in such a plane:  $E^0\mapsto - E^0$, $E^\pm\mapsto E^\mp$. 
As  $f_{\pm}(-s)=f_{\pm}(1\!+\!s)=f_{\mp}(s)$, this is equivalent to 
$\overline{x}^1\mapsto \overline x{}^1$,
$\overline x{}^2\mapsto -\overline x{}^2$, $\overline{L}\mapsto -\overline{L}$. 

\subsection{Convergence to $O(2)$-covariant quantum mechanics on $S$ as $\Lambda\to\infty$}

Define the natural $O(2)$-covariant embedding \  \ ${\cal I}:\Hi_\Lambda\hookrightarrow {\cal L}^2(S)\!\equiv\!\Hi_s$ \ \ by 
  setting \  ${\cal I}\left(\psi_m\right):=u^m$ \ ($u\!\equiv\!e^{i\varphi}$) and applying
linear extension;
below we drop ${\cal I}$  and identify $\psi_m=u^m$ as elements of the Hilbert space.
Clearly $P_\Lambda\phi \to\phi$ in the $\Hi_s$-norm $\Vert\,\Vert$, for all $\phi\in \Hi_s $: \
$\Hi_\Lambda$ `invades' $\Hi_s$ as $\Lambda\to\infty$. 

${\cal I}$ induces an embedding \ \
${\cal J}\!:\!\A_\Lambda\!\hookrightarrow\! B\left[\Hi_s\right]$ \  in the operator algebra,
with $\A_\Lambda$ annihilating $\Hi_\Lambda^\perp$; one easily finds that 
$\overline{L}\!=\!L$ \ on $\Hi_\Lambda$, and \ $\overline{L}\phi\to L\phi$ \ as $\Lambda\to\infty$, \  
for all $\phi\in D(L)\subset\Hi_s$. \
Bounded (resp. continuous) functions $f$ on $S$, acting as multiplication operators 
$f\cdot:\phi\in\Hi_s\mapsto f\phi\in\Hi_s$, make up a subalgebra  $B(S)$ [resp. $C(S)$]
of $B\left[\Hi_s\right]$. \
The fuzzy analog  of  the vector space $B(S)$ is:
\be
{\cal C}_\Lambda:=\left\{\sum_{h=-2\Lambda}^{2\Lambda}f_h \eta^h\:,\: f_h\in\CC\right\},
\label{def_CLambda}
\ee
where $\eta^m\!\equiv\!(\sqrt{2}\xi^+)^m$, $\eta^{-m}\!\equiv\!(\sqrt{2}\xi^-)^{m}$, 
 $m\!\ge\!0$. \ ${\cal C}_\Lambda\subset\A_\Lambda$ as a vector space, but not as a subalgebra.
One easily shows that \ $\eta^h\phi\to u^h\phi$. \ 
Moreover, setting \ $\hat f_\Lambda:=\sum_{h=-2\Lambda}^{2\Lambda}f_h \eta^h\in\A_\Lambda$
for all $f\in B(S)$, we find 

\smallskip
{\bf Proposition 3.3 in \cite{FioPis17}.}  \ Choose $k(\Lambda)\!\ge\! 2 {{\Lambda}}({{\Lambda}}\!+\!1)(2{{\Lambda}}\!+\!1)^2$. Then \ $\hat f_\Lambda\to f\cdot$,  $\widehat{(fg)}_\Lambda\to fg\cdot$, \ $\hat f_\Lambda\hat g_\Lambda\to fg\cdot$ \ strongly as $\Lambda\to\infty$, \ 
$\forall f,g\in B(S)$.

\smallskip
On the other hand, the corresponding convergences in the operator norm do not hold, because
for all $\Lambda\!>\!0$ the operators $\overline{x}^\pm,\overline{L}$ annihilate $\Hi_\Lambda^\perp$, 
whereas  $u^{\pm 1},L$  do not.

\section{D=3:$O\left(3\right)$-covariant fuzzy sphere}
\label{D3}

We associate the pseudovector $L_i=\frac{1}{2}\varepsilon^{ijk}L_{jk}$ to the antisymmetric matrix $L_{ij}$ of the angular momentum components.
For all vectors $\bv$ depending on $\bx,i\nabla$
we shall use either the components $v^i$ ($i\in\{1,2,3\}$) or the ones $v^a$ ($a\in\{-,0,+\}$) defined by
\be
\left(\begin{array}{c} v^+\\ v^-\\ v^0\\
\end{array}\right)  = \: 
\underbrace{\!\!\left(
\begin{array}{ccc}
\frac{1}{\sqrt{2}}&\frac{i}{\sqrt{2}}&0\\
\frac{1}{\sqrt{2}}&\frac{-i}{\sqrt{2}}&0\\
0&0&1\\
\end{array}
\right)\!\!}_{U} \left(\begin{array}{c} v^1\\ v^2\\ v^3\\
\end{array}\right).\label{trasfcoord}
\ee
($U$ is a unitary matrix) which fulfill  
\bea
 [L_a,v^a]=0, \qquad [L_0,v^\pm]=\pm v^\pm, \qquad [L_\pm,v^\mp]=\pm  v^0, \qquad
[L_\pm,v^0]=\mp  v^\pm.          \label{basic_so3'}
\eea
In particular, \ $x^0\!\equiv\!z$, $x^\pm\!=\!\frac{x^1\pm ix^2}{\sqrt{2}}\!\equiv\!\frac{x\pm iy}{\sqrt{2}}\!=\!\frac{r\sin{\theta}e^{\pm i\varphi}}{\sqrt{2}}$. 
We make the Ansatz \ $\psi=\frac{f(r)}{r}Y_l^m\left(\theta,\varphi\right)$. \  $Y_l^m$ are the spherical harmonics:
$$
L^2\, Y_l^m\!\left(\theta,\varphi\right)=l(l+1)Y_l^m\!\left(\theta,\varphi\right),\hspace{1cm}L_3\,Y_l^m\!\left(\theta,\varphi\right)=mY_l^m\!\left(\theta,\varphi\right),
$$ 
with $l\in\mathbb{N}_0$, $m\in\mathbb{Z}, |m|\leq l$. 
Under assumption (\ref{cond1}) the harmonic oscillator approximation of (\ref{eqpolar}) admits 
the 
(H\'ermite) eigenfunctions 
$$
f_{n,l}(r)=N_{n,l} e^{-\frac{\left(r-\widetilde{r}_l\right)^2\sqrt{k_l}}{2}}H_n\left(\left(r-\widetilde{r}_l\right)\sqrt[4]{k_l}\right),
\qquad n=0,1,....
$$
with $k_l\!:=\!2k\!+\!3l(l\!+\!1\!)\!$, $\widetilde{r}_l\!=\!\frac{2k\!+\!4l(l\!+\!1)}{2k\!+\!3l(l\!+\!1)}$. 
\ We  set $V_0\!=\!-\sqrt{\!2k}$ to fulfill the requirement $E_{0,0}=0$; 
then the energies  associated to \ $\psi_{n,l,m}=\frac{f_{n,l}(r)}{r}Y_l^m\!\left(\theta,\!\varphi\right)$ \ are
$$
E_{n,l}={\color{red}{2n\sqrt{2k}}}+{\color{blue}{l(l+1)}}+O\left(1/\sqrt{2k}\right).
$$
Again $E_{0,l}={\color{blue}{l(l+1)=: E_l}}$ are the eigenvalues of the Laplacian  $L^2$ on $S^2$, while $E_{n,l}\to\infty$ as $k\!\to\!\infty$ if $n\!>\!0$  (we have highlighted the unwelcome $n$-dependent contribution  in red,  the welcome $l$-dependent one in blue). 
We can eliminate the latter (i.e. constrain $n=0$) imposing a cutoff 
\begin{equation}
 E\le \Lambda(\Lambda+1)\equiv\overline{E} <2\sqrt{\!2k}, \label{consistency3D}
\end{equation}
namely  projecting the theory on the Hilbert  subspace
\ $\mathcal{H}_{\Lambda}\!\subset\!
{\cal L}^2(\RR^3)$  spanned by 
\be
\psi_l^m:=\psi_{0,l,m}\simeq\frac{N_l}{r}e^{-\frac{\left(r-\widetilde{r}_l\right)^2\sqrt{k_l}}{2}}\, Y_l^m(\theta,\varphi), \quad |m|\leq l, \quad l\leq\Lambda. 
\ee
Clearly \  dim$(\Hi_{\Lambda})\!=\!(\Lambda\!+\!1)^2$. \ \ Multiplication by \ $x^a=r\frac{x^a}{r}$ \ 
($a=-,0,+$) on $\psi_l^m$ factorizes into the one by $r$ on $\frac{f_{0,l}(r)}{r}$ and the one
 by $\frac{x^a}{r}$ on $Y_l^m$. After projection we find
\smallskip
\bea
\ba{l}
\overline{x}^{a}\psi_l^m=c_lA_l^{a,m}\psi_{l-1}^{m+a}+c_{l+1} 
A^{-a,m+a}_{l+1}\psi_{l+1}^{m+a}, \\[10pt] 
c_0=c_{\Lambda+1}=0, \qquad 
c_l= \sqrt{1+\frac{l^2}{k}}\quad 1\le l\le \Lambda
\ea\label{3Daction}
\eea
up to terms $O\left(1/k^{\frac{3}{2}} \right)$; \ here $A^{a,m}_l ,B_l^{a,m}$ are the coefficients determined by
the ordinary multiplication rules
$$
\frac{x^{a}}r Y^m_l=A^{a,m}_l Y^{m+a}_{l-1}+B^{a,m}_l
Y^{m+a}_{l+1},
$$
namely $B_l^{a,m}=A_{l+1}^{-a,m+a}$, and
\bea
A^{\pm,m}_l =\pm\sqrt{\frac{(l\!\mp\!m)(l\!\mp\!m\!-\!1)}{2(2l\!+\!1)(2l\!-\!1)}},\qquad
A^{0,m}_l =\sqrt{\frac{(l+m)(l-m)}{(2l+1)(2l-1)}}.
\eea
At leading order the $\overline{L}_i,\overline{x}^i$, $i\in\{1,2,3\}$,   fulfill 
\bea
&& \prod_{l=0}^{\Lambda}\left[\overline{L}^2-l(l+1)I\right] =0,\qquad
\prod_{m=-l}^{l}{\left(\overline{L}_3-mI\right)}\widetilde{P}_l=0, \label{rf3D3}\\[4pt]
&& \overline{L}_i ^{\dag}=\overline{L}_i,  \quad \left[\,\overline{L}_i,\overline{L}_j\right]=i\varepsilon^{ijh}\overline{L}_h,\qquad \overline{x}^{i\dag}=\overline{x}^i, \quad \overline{x}^i\overline{L}_i=0,\label{rf3D4}\qquad\\[8pt]
&& {\color{blue}\underbrace{[\overline{L}_i,\overline{x}^j]=i\varepsilon^{ijh}\overline{x}^h}_{Snyder-like}},\hspace{0.9cm}
[\overline{x}^i,\overline{x}^j]=i\varepsilon^{ijh}\underbrace{\left(-{\color{blue}\frac{1}{k}}+K\widetilde{P}_{\Lambda}\right){\color{blue}\overline{L}_h},}_{{\color{blue}Snyder-like}}\label{xx}
\eea
where $K=\frac{1}{k}\!+\!\frac{1\!+\!\frac{\Lambda^2}{k}}{2\Lambda+1}$, \ $\overline{L}^2:=\overline{L}_i\overline{L}_i=\overline{L}_a\overline{L}_{-a}$ is  $L^2$ projected on $\Hi_\Lambda$, and $\widetilde{P}_l$ is the projection on its eigenspace  with eigenvalue $l(l+1)$. 
Moreover, the square distance from the origin is the $O(3)$-invariant 
\be
\mathcal{R}^2:=\overline{x}^{i}\overline{x}^{i}= 1
+\frac{\overline{L}^2+1}{k}-\left[1+\frac{(\Lambda\!+\!1)^2}{k}\right]\frac{\Lambda\!+\!1}{2\Lambda+1}\widetilde P_{\Lambda}.
\label{R^2D=3}
\ee
Relations (\ref{rf3D3}-\ref{R^2D=3}) are {\it exact} if we adopt (\ref{3Daction}) as {\it definitions} of $\overline{x}^a$.
To obtain a fuzzy space we can choose $k$ as a function of $\Lambda$ fulfilling (\ref{consistency3D}); one possible choice is $k=\Lambda^2(\Lambda+1)^2$, and the commutative limit will be $\Lambda\rightarrow+\infty$.
Again: 

\begin{itemize} 

\item The commutators $[\overline{x}^i,\overline{x}^j]$ (\ref{xx}) depend only on the angular momentum and
 are Snyder-like, i.e.  (apart from the additional term in the second formula) are proportional to angular momentum components, and vanish  as \ $\Lambda\!\to\! \infty$; \ in the same limit $\psi_l^m\!\to\!\delta(r\!-\!1)Y_l^m$.   

\item  Hence (\ref{rf3D3}-\ref{xx})  are covariant under the whole group $O(3)$, including
parity $\overline{x}_i\!\mapsto\!-\overline{x}_i$, $\overline{L}_i\!\mapsto\!\overline{L}_i$,
contrary to Madore's and Hoppe's FS.

\item  $\mathcal{R}^2\neq 1$; but it is a function of $L^2$, and,  for each fixed $\Lambda$, its eigenvalues  (except the highest one)
are close to 1, slightly grow with $l$ and collapse to 1 as $\Lambda\to \infty$.

\item  The ordered monomials in $\overline{x_i},\overline{L_i}$ 
make up a 
 basis of the $(\Lambda\!+\!1)^4$-dim  vector space \ $\A\!:=\!End(\Hi_\Lambda\!)\!\simeq\! M_{(\Lambda+1)^2}(\CC)$, 
because the $\widetilde{P}_l$ themselves can be expressed as polynomials in $\overline{L}^2$.

\item Actually,  $\overline{x}_i$ {\it generate} the $*$-algebra $\A$, because
also the $\overline{L}_i$ can be expressed as a non-ordered polynomial in the $\overline{x}_i$.

\end{itemize}

\subsection{Realization of the algebra $\A_\Lambda$ of observables through $Uso(4)$}

We recall that \ $so(4)\simeq su(2)\oplus su(2)$; \ hence this Lie algebra is spanned by
$\left\{E_i^1,E_i^2\right\}_{i=1}^3$ fulfilling
\bea
[E^1_i, E^2_j]=0,\qquad [E^1_i, E^1_j]=i\varepsilon^{ijk}E^1_k,\qquad [E^2_i, E^2_j]=i\varepsilon^{ijk}E^2_k.
\label{CRE}
\eea
$L_i:=E_i^1+E_i^2$, \  $X_i:=E_i^1-E_i^2$ make up  alternative basis  of $so(4)$ and fulfill
\bea
[L_i, L_j]=i\varepsilon^{ijk}L_k, \qquad[L_i, X_j]=i\varepsilon^{ijk}X_k,\qquad [X_i, X_j]=i\varepsilon^{ijk}L_k.
\label{CRXL}
\eea
The $L_i$ close another $su(2)$.  Passing to generators labelled by $a\in\{-,0,+\}$, we find
\bea
\left[L_+,L_-\right]=L_0,\quad \left[L_0,L_{\pm}\right]=\pm L_{\pm}=[X_0,X_\pm],\quad [X_+,X_-]= L_0, \label{u1}\\[6pt]
 [L_\pm,X_\mp ]=\pm X_0,\quad [L_0,X_\pm]=\pm X_\pm= [X_0,L_\pm],
\quad   [L_a,X_a]=0 \label{u2}
\eea
(in the last formula there is no sum over $a$), where  \ $L^2\!:= L_iL_i=L_aL_{-a}$, $X^2\!:=X_iX_i=X_aX_{-a}$. 

In the representation $\bpi_{\Lambda}:=\pi_{\frac{\Lambda}{2}}\otimes  \pi_{\frac{\Lambda}{2}}$\  of $Uso(4)\simeq U\!su(2)\otimes U\!su(2)$ \ on the Hilbert space 
${\bf V}_{\Lambda}:=V_{\frac{\Lambda}{2}}\otimes V_{\frac{\Lambda}{2}}$ it is $C^1:=E^1_iE^1_i=\frac{\Lambda}{2}(\frac{\Lambda}{2}+1)=E^2_iE^2_i=:C^2$, or equivalently
\be
X\cdot L=L\cdot X=0,\qquad X^2\!+\!L^2=\Lambda(\Lambda\!+\!2)\label{u3}
\ee
(we have  dropped the symbols $\bpi_{\Lambda}$).  ${\bf V}_{\Lambda}$ 
admits an orthonormal basis consisting of common eigenvectors of $L^2$ and $L_0$; in standard ket notation,
\be
L_0\left\vert l,m\rangle\right.=m\left\vert l,m\rangle\right., \qquad L^2\left\vert l,m\rangle\right.=l(l+1)\left\vert l,m\rangle\right.
\ee
with $ 0\leq l\leq \Lambda\mbox{ and }|m|\leq l$. \ ${\bf V}_{\Lambda},\Hi_{\Lambda}$ have
the same dimension $(\Lambda\!+\!1)^2$ and  decomposition in irreducible representations
of the $L_i$ subalgebra; we identify them  setting $\psi_l^m\equiv \vert l,m\rangle$. \ The action of $X^a$ on ${\bf V}_{\Lambda}$ reads
\bea
&& X^a\left\vert l,m\rangle\right.=d_l A_l^{a,m}\left\vert l-1,m+a\rangle\right.+d_{l+1}B_l^{a,m}\left\vert l+1,m+a\rangle\right. \qquad\\
&& d_{l}:=\sqrt{(\Lambda\!+\!1)^2-l^2} \nonumber
\eea

We can naturally  realize \ $\overline{L}_a,\, \overline{x}^a$ \ within \ ${\bf{\pi}}_{\Lambda}\left[U\!su(2)\otimes Usu(2)\right]$
\cite{FioPis17}. 
Define $\lambda:=\frac{\sqrt{4L^2+1}-1}2$; \ then $\lambda\,\vert l,m\rangle=l\,\vert l,m\rangle$. \ The Ansatz
\be
\overline{L}_a=L_a,\qquad \overline{x}^a=g(\lambda)\, X^a\,g(\lambda), \label{repr}
\ee
fulfills (\ref{3Daction}) and therefore (\ref{rf3D3}-\ref{xx}), provided
\bea
g(l) &=& \sqrt{\frac{\prod_{h=0}^{l-1}(\Lambda\!+\!l\!-\!2h)}{\prod_{h=0}^l(\Lambda\!+\!l\!+\!1\!-\!2h)}
\prod_{j=0}^{\left[\frac{l\!-\!1}2\right]}\frac{1+\frac{(l\!-\!2j)^2}k}{1+\frac{(l\!-\!1\!-\!2j)^2}k}} \\
 &=& \sqrt{
\frac{\Gamma\!\left(\frac {\Lambda\!+\!l}2\!+\!1\right)\Gamma\!\left(\frac {\Lambda\!-\!l\!+\!1}2\right)}
{\Gamma\!\left(\frac {\Lambda\!+\!1\!+\!l}2\!+\!1\right)\Gamma\!\left(\frac {\Lambda\!-\!l}2\!+\!1\right)}
\frac{\Gamma\!\left(\frac l2\!+\!1\!+\!\frac{i\sqrt{k}}2\right)\Gamma\!\left(\frac l2\!+\!1\!-\!\frac{i\sqrt{k}}2\right)}
{\sqrt{k}\:\Gamma\!\left(\frac {l\!+\!1}2\!+\!\frac{i\sqrt{k}}2\right)\Gamma\!\left(\frac {l\!+\!1}2\!-\!\frac{i\sqrt{k}}2\right)}};\nonumber
\eea
here we have used Euler's gamma-function $\Gamma$.
The inverse of  (\ref{repr}) is clearly $X^a=[g(\lambda)]^{-1}\,\overline{x}^a\,[g(\lambda)]^{-1}.$
We have thus explicitly constructed a *-algebra isomorphism
\be
\A_\Lambda:=End(\Hi_{{{\Lambda}}})\simeq M_N(\CC)\simeq\pi_\Lambda[Uso(4)], \quad N:=(\Lambda\!+\!1)^2.             \label{isomD4}
\ee

\subsection{$*$-Automorphisms of the algebra of observables}

Within the group of  $*$-automorphisms of 
the algebra of observables $M_N(\CC)\simeq\A_\Lambda$
$$
b\rightarrow gbg^{-1}, \qquad b\in\mathcal{A}_\Lambda, \quad g\in SU(N)
$$ 
again a special role is played 
by the subgroup $SO(4)$ acting through the representation $\bpi_{\Lambda}$, namely $g=\bpi_{\Lambda}\left[e^{i\alpha}\right]$, $\alpha\in so(4)$.  
$O(3)\subset SO(4)$ plays the role of  isometry subgroup.
In particular, choosing $\alpha=\alpha_iL_i $ ($\alpha_i\in\RR$)  the automorphism amounts to 
  a $SO(3)$ transformation (a rotation in $\RR^3$). 
An $O(3)$ transformation
with determinant  $-1$ in the $X^1X^2X^3$ space, and therefore also  
in the $ \overline{x}^1 \overline{x}^2 \overline{x}^3$ space, is parity: \ $(L_i,X^i)\mapsto (L_i,-X^i)$, or equivalently
$E_i^1\leftrightarrow E_i^2$, the only automorphism
of $so(4)$ (corresponding to the exchange of the two nodes in the Dynkin diagram).

\subsection{Convergence to $O(3)$-covariant quantum mechanics on $S^2$ as $\Lambda\to\infty$}

Define the $O(3)$-covariant embedding \  \ ${\cal I}:\Hi_\Lambda\hookrightarrow {\cal L}^2(S^2)\!\equiv\!\Hi_s$ \ \ by 
 setting \  ${\cal I}\left(\psi_l^m\right):=Y_l^m$ \
and applying linear extension; \ \
below we drop ${\cal I}$  and identify $\psi_l^m=Y_l^m$  as elements of the Hilbert space.
Clearly $P_\Lambda\phi \to\phi$ in the $\Hi_s$-norm $\Vert\,\Vert$, for all $\phi\!\in\!\Hi_s$; \ 
$\Hi_\Lambda$ `invades' $\Hi_s$ as $\Lambda\to\infty$. 

${\cal I}$ induces the embedding of operator algebras \ \
${\cal J}\!:\!\A_\Lambda\!\hookrightarrow\! B\left[\Hi_s\right]$,
with $\A_\Lambda$ annihilating $\Hi_\Lambda^\perp$; \ 
$\overline{L_i}\!=\!L_i$ \ on $\Hi_\Lambda$, and \ $\overline{L}_i\phi\to L_i\phi$ \ as $\Lambda\to\infty$, \  
 for all $\phi\!\in\! D(L_i)\subset\Hi_s$.
Bounded (resp. continuous) functions $f$ on $S^2$, acting as multiplication operators 
$f\cdot:\phi\in\Hi_s\mapsto f\phi\in\Hi_s$, make up a subalgebra  $B(S^2)$ [resp. $C(S^2)$]
of $B\left[\Hi_s\right]$. \
We define the fuzzy analog  of  the vector space $B(S^2)$ as
\be
{\cal C}_\Lambda:=\left\{\sum_{l=0}^{2\Lambda}\sum_{m=- l}^l
f_l^m \widehat{Y}_l^m\:,\: f_l^m\in\CC\right\},
\label{def_CLambda3D}
\ee
\be
\mbox{where}\qquad\widehat{Y}_l^m:=M_l\sqrt{\frac{(l+m)!2^{l-m}}{(2l)!(l-m)!}} \:
L_-^{l-m}(\overline{x}^+)^l                  \label{defhatY}
\ee
are the fuzzy analogs of 
\  $Y_l^m\cdot \in B(S^2)$. \  ${\cal C}_\Lambda\subset\A_\Lambda$ as a vector space, but not as a subalgebra.
The decomposition of  ${\cal C}_\Lambda$ in irreducible representations of $O(3)$ reads \ 
 ${\cal C}_\Lambda=\bigoplus_{l=0}^{2\Lambda} V_l$.
In  \cite{FioPis17} we have shown that for all $\phi\!\in\!\Hi_s$ \ $\overline{x}^i\phi\to (x^i/r)\phi$; \ 
more generally,  setting  \ $\hat f_\Lambda:=\sum_{l=0}^{2\Lambda}\sum_{|m|\leq l}f_l^m 
\widehat{Y}_l^m\in\A_\Lambda$ \ for all
$f\in B(S^2)$ \ we find 

\smallskip
{\bf Proposition 4.3 in \cite{FioPis17}.} \ Choose $k(\Lambda)\ge 2^{3\Lambda+3}\Lambda ^{\Lambda+5}(\Lambda\!+\!1)$. Then \ $\hat f_\Lambda\to f\cdot$,  $\widehat{(fg)}_\Lambda\to fg\cdot$, \ $\hat f_\Lambda\hat g_\Lambda\to fg\cdot$ \ strongly as $\Lambda\to\infty$, \ 
$\forall f,g\in B(S^2)$.

\smallskip
On the other hand, the corresponding convergences in the operator norm do not hold, because
for all $\Lambda\!>\!0$ the operators $\overline{x}^i,\overline{L}_i$ annihilate $\Hi_\Lambda^\perp$, 
whereas  $x^i/r,L_i$  do not.

\section{Comparison with the literature, final remarks and outlook}
\label{Conclu}

In conclusion, for \ $d=1,2$ \  we have built  a sequence $(\A_\Lambda, \Hi_\Lambda)$
of finite-dim, $O(D)$-covariant  ($D=d\!+\!1$)  approximations of quantum mechanics of a spinless
particle on the sphere  $S^d$; \ \ ${\cal R}^2\gtrsim 1$ collapses to 1 as $\Lambda\to\infty$. \
This result has been achieved imposing  an energy-cutoff 
$E\le \Lambda(\Lambda\!+\!d\!-\!1)$ on quantum mechanics of  a  particle in $\RR^{D}$  subject to a sharp confining potential $V(r)$  on the  sphere $r=1$. 
$\A_\Lambda$ is a fuzzy approximation of the {\it whole algebra of observables} of  the  particle on $S^d$
(phase space algebra), and converges to it in the limit
$\Lambda\to \infty$. We have explicitly determined a $*$-isomorphism 
$\A_\Lambda\simeq \pi_\Lambda[Uso(D\!+\!1)]$, 
with a suitable irreducible representation  $\pi_\Lambda$ of $U\!so(D\!+\!1)$ on $\Hi_{\Lambda}$. 
On the other hand $\Hi_{\Lambda}$ carries 
a {\it reducible} representation of the $U\!so(D)$ subalgebra 
generated by the $\overline{L}_{ij}$: \ \
 $\Hi_{\Lambda}$ is the direct sum of {\it all} irreducible representations fulfilling  $L^2\le \Lambda(\Lambda\!+\!d\!-\!1)$.
A similar decomposition holds for the subspace ${\cal C}_\Lambda\subset\A_\Lambda$ 
of completely symmetrized polynomials in the $\overline{x}^i$ acting as multiplication operators on $\Hi_\Lambda$. 
As $\Lambda\to\infty$ 
these respectively become the decompositions (\ref{directsum}) of  ${\cal L}^2(S^d)$ and of 
 $C(S^d)$ acting on ${\cal L}^2(S^d)$.

\smallskip
Our approach seems applicable  to $d\ge 3$; this will allow a more direct comparison with 
the rest of the literature. 
The fuzzy spheres of dimension $d=4$ introduced in \cite{GroKliPre96}, as well as the 
$d\ge 3$ ones considered in \cite{Ramgoolam,DolOCon03,DolOConPre03}, are  based on $End(V)$, 
where $V$ carries a particular {\it irreducible representation} of $SO(D)$; as $\R^2$ is central, it can be set $\R^2=1$ identically. The commutation relations are also Snyder-like,  hence $O(D)$-covariant. 
The fuzzy spherical harmonics are elements of  $End(V)$, but do do not close a subalgebra of $End(V)$,
i.e. the product $Y\cdot Y'$ of two spherical harmonics is not a combination of spherical harmonics.
This is exactly as in our models, i.e.  ${\cal C}_\Lambda$ is a subspace, but not a subalgebra, of
$\A_\Lambda$. (One can introduce a product in ${\cal C}_\Lambda $ by projecting the result of $Y\cdot Y'$ 
to the vector space ${\cal C}_\Lambda$, but it will be non-associative).

In \cite{Ste16,Ste17} Steinacker and Sperling consider the possibility of construncting a fuzzy 4-sphere $S^4_N$
through a {\it reducible} representation of $Uso(5)$ on a Hilbert space $V$ obtained decomposing
 an irreducible representation $\pi$ of $Uso(6)$ characterized by a triple of highest weights
$(N,n_1,n_2)$; so $End(V)\simeq  \pi[Uso(6)]$, in analogy with our result. 
The elements $X^i$ of a basis of the vector space \ $so(6)\setminus so(5)$ \
play the role of noncommuting cartesian coordinates.  
Hence, the $O(5)$-scalar $\R^2=X^iX^i$ is no longer central, but its spectrum is still very close to 1  {\it provided $N\gg n_1,n_2$}, because then $V$ decomposes only in few irreducible $SO(5)$-components, all with
eigenvalues of $\R^2$ very close to 1; 
if $n_1=n_2=0$ then $\R^2\equiv 1$ ($V$ carries an irreducible representation of $O(5)$), and one recovers the  fuzzy 4-sphere
of   \cite{GroKliPre96}. On the contrary, in our approach 
 $\R^2\simeq 1$  is guaranteed  by adopting
 $\overline{x}^i=g(L^2)X^ig(L^2)$ rather than $X^i$ as noncommutative cartesian coordinates, and \
$\R^2=\overline{x}^i\overline{x}^i$. 

\smallskip
Many other aspects of these new fuzzy spheres deserve  investigations: e.g.   
space uncertainties, optimally localized states (coherent states \cite{Perelomov}), their distance
(as done e.g. in \cite{DanLizMar14} for the FS), extension to particles with spin, etc.
We hope that progresses on these and other issues can be reported soon.

\subsubsection*{Acknowledgments}

We are grateful to F. D'Andrea and T. Weber for useful discussions.
This article is based upon work from COST Action MP1405 QSPACE (Quantum Structure of Spacetime), 
supported by COST (European Cooperation in Science and Technology).

\end{document}